# Substrate induced optimization of the Electrocatalytic Hydrogen Evolution Reaction (HER) performances of MoS$_2$ thin films

Hafiz Sami-Ur-Rehman,*[a] Arpana Singh,[a] Nunzia Coppola,[a] Pierpaolo Polverino,[a] Sandeep Kumar Chaluvadi,[b] Shyni Punathum-Chalil,[bc] Heinrich-Christoph Neitzert,[a] Diana Sannino,[a] Pasquale Orgiani,[bd] Alice Galdi,[a] Cesare Pianese,[a] Paolo Barone,[e] Carmela Aruta,[e] Luigi Maritato[a]

[a] *Dipartimento di Ingegneria Industriale-DIIN, Università Degli Studi di Salerno, 84084 Fisciano (SA), Italy.*
[b] *CNR-IOM, Strada Statale 14 Km 163.5, 34149 Basovizza, Trieste, Italy.*
[c] *International Centre for Theoretical Physics (ICTP), Str. Costiera 11, I-34151 Trieste, Italy.*
[d] *AREA Science Park, Padriciano 99, I-34149 Trieste, Italy.*
[e] *CNR-SPIN, Via del Fosso del Cavaliere 100, 00133, Roma, Italy.*
* Email: hsamiurrehman@unisa.it

**Abstract**

Molybdenum disulfide (MoS$_2$) has emerged as a promising, cost-effective catalyst for hydrogen production via water splitting. We investigate the structural and electrocatalytic properties of MoS$_2$ thin films deposited on different substrates (Al$_2$O$_3$, SiC, STO) to study their hydrogen evolution reaction (HER) activity. In particular, in order to study the substrate influence on the stabilization of different polymorphic MoS$_2$ phases, the films are synthesised using pulsed laser deposition on substrates with different crystal symmetries and lattice parameters. All the deposited samples are characterized by X-Ray Diffraction, Raman Spectroscopy, Linear Sweep Voltammetry and Electrochemical Impedance Spectroscopy analyses. The films grown on Al$_2$O$_3$ substrates exhibit the best HER performance, likely due to the stabilization of the metastable 1T phase through the interfacial interactions between film and substrate. Presence of the 1T phase in the samples grown on Al$_2$O$_3$ improves the charge transfer efficiency and the electrochemically active surface with a better response to the applied potential, demonstrating their enhanced catalytic behaviour for hydrogen evolution.

**Introduction**

The growing global energy demands, and the depletion of fossil fuel resources have drawn significant attention to the development of alternative, sustainable and environmentally friendly renewable resources[1]. The excessive use of fossil fuels such as coal, oil and natural gas, intensified the CO$_2$ emissions that significantly contributed to climate change, posing a serious threat to human life. In the quest for clean energy, hydrogen has emerged as a promising fuel for a sustainable future, able to mitigate the greenhouse effect due to in principle zero carbon emission and three times higher energy density (gravimetric chemical energy up to 142 MJ Kg$^{-1}$) compared to fossil fuels[2]. However, at the present, the most largely used industrial hydrogen production methods are inevitably accompanied by CO$_2$ emissions.

Among the various H$_2$ production methods, water splitting is considered the most eco-friendly and can be accomplished through electrochemical, photochemical or photoelectrochemical techniques. For the electrochemical large-scale hydrogen production, the hydrogen evolution reaction (HER) is a fundamental process for which platinum supported on carbon (Pt/C) catalysts are usually employed[3,4]. The main challenge in the industrial implementation of large-scale electrochemical water splitting is the sluggish reaction kinetics which require large overpotential for hydrogen generation[5]. To overcome these challenges, efficient and inexpensive electrocatalysts are essential to achieve economically viable large-scale water splitting.

Therefore, the development of efficient, stable, cost-effective electrocatalysts from abundant materials represents a critical issue. Due to the high cost and scarcity of the noble metals, extensive research has been directed towards non-precious metal-based catalysts such as transition metal chalcogenides (TMDs), phosphides, carbides, borides, functional metal-free carbon and various composites[6–8].

Over the past few decades, molybdenum disulfide ($MoS_2$), with its S-Mo-S triatomic layer structure, has emerged as one of the mostly explored TMDs catalytic materials, due to its hydrogen absorption energy being comparable to that of Pt[9,10]. $MoS_2$ exhibits four polymorphic crystal structures: 1H, 1T, 2H and 3R[11]. The coordination geometry of Mo atoms and the stacking sequence of the S-Mo-S layer define the distinct polymorphic phases. The most studied are the 2H and 1T phases, consisting of hexagonal and tetragonal coordination of Mo atoms and exhibiting semiconducting (2H) and metallic (1T) behaviour, respectively. The 1T-$MoS_2$ phase is thermodynamically metastable and easily convertible into the 2H-$MoS_2$ by regulating the electron occupancy in the d orbitals[12–14]. Regarding the electrocatalytic properties of the 2H phase, Jaramillo et al. found that the active sites for HER are primarily located at the edges of $MoS_2$ rather than on the basal plane[15]. Since the number of edge sites is limited, much research has focused on activating the basal plane through phase engineering[16,17], defect engineering[18,19], doping[20–22] and strain engineering[23,24]. These strategies have proven to significantly enhance the HER performances[19]. However, $MoS_2$ studies are mainly based on the 2H phase, and only recent investigations report that the 1T phase may have better catalytic efficiency for HER[25–27]. This improved activity arises from the presence of catalytically active sites not only at the edges, as in the 2H phase, but also across the basal planes, thereby significantly increasing the site density. In addition, the 1T phase exhibits metallic conductivity due to its electronic structure, which facilitates charge transfer and enhances the kinetics of the electrochemical reaction. The role of substrates in modulating the 1T-phase presence in the $MoS_2$/hBN thin films has been demonstrated[28]. The atomically thin nature of the 2D $MoS_2$ film suggests that substrate effects are expected to be substantial and recent studies have indeed shown that the substrate strongly influences other properties such as band structure and optical properties[29,30]. Theoretical studies of the catalytic activity of the 2D $MoS_2$ film on metallic substrates have addressed the influence of such substrates on the edge sites available for HER[31,32].

However, the current understanding of the interplay between film and substrate is generally poor and limited to provide instructive guidance to optimise HER catalytic performance. Here, with the aim to elucidate further the film/substrate mechanisms and to optimise the final electrochemical and HER catalytic performances of the of $MoS_2$, we have investigated the structural and electrochemical properties of $MoS_2$ films deposited on different substrates ($SrTiO_3$ (111), *c*-axis $Al_2O_3$ (0001), 6H-SiC (0001)) with different crystal symmetries and lattice parameters. The interaction between film and substrate can strongly influence HER kinetics by promoting the stabilization of different $MoS_2$ polymorphs. The Pulsed Laser Deposition (PLD) grown $MoS_2$ films have been characterized by X-Ray Diffraction (XRD), Raman Spectroscopy, Linear Sweep Voltammetry (LSV) and Electrochemical Impedance Spectroscopy (EIS) measurements. Our results suggest that the presence of different substrate-film interaction plays an important role in determining and stabilizing the fraction of the 1T-phase. Moreover, the LSV and EIS analyses indicate that the Tafel slopes for all the films were found to be relatively consistent across the different substrates, with values typically around 75 mV dec$^{-1}$ and that the best catalytic properties of $MoS_2$, with j=10 mA/cm$^2$ at -0.2 V overpotential, are obtained in the case of sample deposited on $Al_2O_3$, because of the optimization between the electrode/electrolyte charge transfer and the number of active sites.

**Results and discussion**

XRD measurements in symmetric configuration of the $MoS_2$ films grown on three different substrates are shown in Figure 1. All the films exhibit a distinct diffraction peak of $MoS_2$ located at 2θ ≈ 14.20°, 14.04° and 13.86°, for $Al_2O_3$, STO and SiC, respectively. These values are very close to the 2H-$MoS_2$ (002) and the 1T-$MoS_2$ (001) reflections, corresponding to 14.37° for the first and 14.89° for the second phase. Indeed, 2H-$MoS_2$ has a hexagonal crystal structure with a=3.169 Å and c=12.324 Å, while 1T-$MoS_2$ has trigonal crystal structure with a=3.190 Å and c=5.945 Å[33,34]. The absence of any additional diffraction peaks associated with other $MoS_2$ orientations indicates that the films are predominantly *c*-axis oriented, with the basal planes aligned parallel to the substrate surface[35,36]. Moreover, no diffraction peaks were detected in asymmetric configuration, demonstrating that the films are not epitaxial, but more likely develop domains with different in-plane orientations. The shift of the diffraction peaks in panel b) indicates that the *c*-axis of the films varies depending on the substrate.

Due to the large differences in the substrate crystal symmetries and lattice parameters ((111) STO cubic with a=3.905 Å, *c*-plane $Al_2O_3$ hexagonal with *a*=4.78 Å and *c*=12.99 Å, 6H-SiC hexagonal with *a*=3.08 Å and *c*=15.117 Å), one could expect an effect on the *c* lattice parameters depending on the in-plane lattice distortion. However, as explained in the Supplementary Information, the observed behaviour of the *c* lattice parameter cannot be exclusively explained in terms of strain effects induced by the substrate.

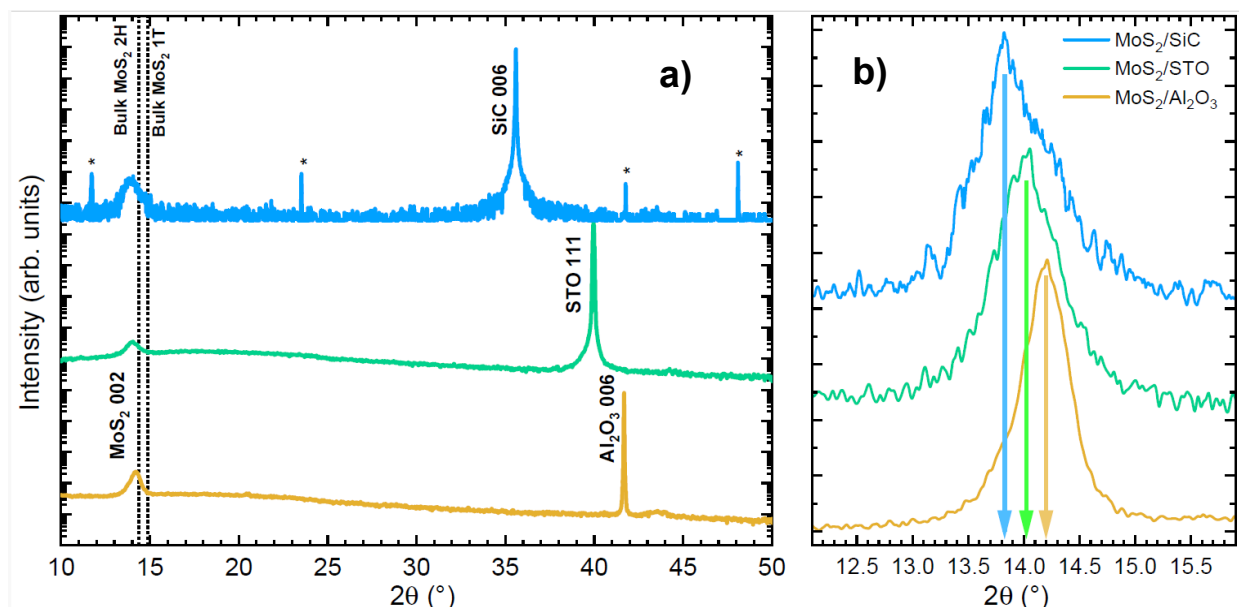

Figure 1. (left) The long-range θ-2θ (10° - 100°) XRD spectra of $MoS_2$ films on different substrates (SiC, STO, and $Al_2O_3$), showing preferential *c*-axis orientation. Dotted lines indicate the positions of the 1T and 2H $MoS_2$ bulk phases. (Right) Magnified view of the 002 $MoS_2$ reflection, with arrows indicating the peak centres.

The systematic increase of the *c*-axis parameter with respect to bulk 2H- and 1T-MoS$_2$ can be attributed to substrate-induced effects on stacking correlations, due to the symmetry, chemical termination, and surface properties. The *c*-plane Al$_2$O$_3$ substrate provides a well-defined in-plane surface which can favour a more correlated stacking sequence, resulting in a smaller effective expansion of the interlayer spacing. On the contrary, STO (111) exhibits a pseudo-hexagonal but non-ideal surface plane with a less defined lateral registry, and 6H-SiC (0001) exposes a covalent and chemically inert hexagonal surface that provides limited stacking selectivity, both resulting in weakly correlated layer stacking selectivity, giving rise to a larger average interlayer distance.

To better analyze the structure of our films, we have, therefore, performed Raman Spectroscopy measurements. The Raman spectroscopy is a powerful and versatile technique able to determine the presence of specific material and phases through their distinct vibrational fingerprints. It provides insight into crystalline phases, layer thickness, interlayer coupling, strain and crystal symmetry[28].

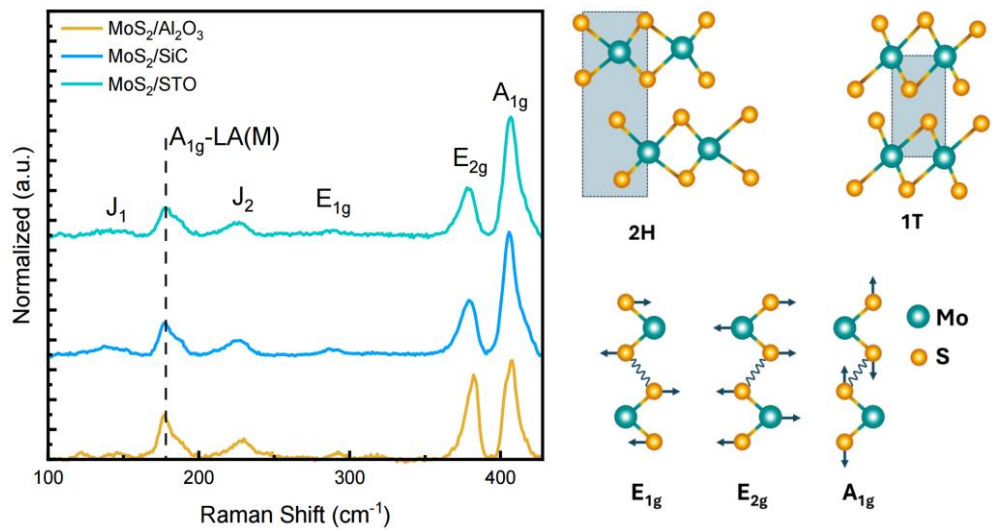

Figure 2. Raman analysis of MoS$_2$ films on different substrates and illustration of the Raman phonon interactions on MoS$_2$ atoms.

In Figure 2, the Raman spectra obtained for the MoS$_2$ films on different substrates are shown. In this figure, the J peaks are associated to vibrational modes in the 1T-phase, while the E and A peaks are related to 2H-phase vibrational modes. In particular, see the sketch in Figure 2, the E$_{2g}$ mode corresponds to the in-plane vibration of Mo and S atoms in opposite direction, while A$_{1g}$ mode corresponds to the out of plane vibration of the S atoms[37–39]. A low intensity peak associated to the E$_{1g}$ mode, corresponding to asymmetric vibration of S atoms relative to Mo atom in the plane of single layer, appears in the spectra of Figure 2, at 287.4 cm$^{-1}$ for MoS$_2$ on SiC, at 289.2 cm$^{-1}$ for STO and at 292.3 cm$^{-1}$ for Al$_2$O$_3$[37,38]. The Raman features J$_1$ can be assigned to the atomic vibrations of the Mo zigzag chain[39–41], while J$_2$ arises from the vibration of the S atoms [28,42,43]. It is interesting to point out that in all the investigated samples we observe the simultaneous presence of the 1T and 2H phases. Contrary to S. Parmar and coworkers[28], we cannot explain our results in terms of the substrate induced strain. However, there may be several other effects that determine the presence of the 1T and 2H phases, such as symmetry, chemical termination and electronic environment at the substrate surface. The occurrence of both the 1T and 2H phases is consistent with our previous findings reported in [21].

In Figure 2, the intensities of all the peaks have been normalized to that of the $A_{1g}$, after subtraction of the $Al_2O_3$ substrate peak contribution (≈ 417 cm$^{-1}$). The Raman peak observed near 179 cm$^{-1}$ is attributed to resonant Raman effects in bulk $MoS_2$[44], which are typically not visible when using 514 or 532 nm excitation, as mostly used in literature[44]. This peak arises from a difference combination mode, $A_{1g}$ – LA(M). The LA(M) mode corresponds to a quasi-longitudinal acoustic vibration at the M point of the Brillouin zone and appears at 233 cm$^{-1}$[44].

In Table 1, we report the values of the Raman shift associated to all the peaks observed in the samples on different substrates. It is interesting to note that the behaviour of all the 2H-peaks is the same, with the lowest shift always observed for the sample on SiC and the highest for the film on $Al_2O_3$. The percentage difference in the peak position [($\omega_{max}$ – $\omega_{min}$)/$\omega_{min}$] observed for the $A_{1g}$ peak is 4.2% and that of the $E_{2g}$ peak is 8.7%, possibly indicating a larger effect induced by substrate interface for the in-plane vibrating modes. The same behaviour is not observed for the J peaks associated to the 1T phase, where the percentage difference in the peak position is smaller.

Table 1. Raman shifts values of the peaks observed in the spectra of Figure 2.

| $MoS_2$/Substrate | $J_1$ (cm$^{-1}$) | $A_{1g}$–LA(M) (cm$^{-1}$) | $J_2$ (cm$^{-1}$) | $E_{1g}$ (cm$^{-1}$) | $E_{2g}$ (cm$^{-1}$) | | $A_{1g}$ (cm$^{-1}$) | |
|---|---|---|---|---|---|---|---|---|
| | | | | | Position | FWHM | Position | FWHM |
| $Al_2O_3$ | 141.3 | 176.9 | 227.3 | 292.3 | 382.2 | 6.9 | 407.4 | 7.1 |
| SiC | 141.3 | 176.9 | 228.0 | 287.4 | 378.9 | 8.8 | 405.7 | 8.3 |
| STO | 141.2 | 179.2 | 226.7 | 289.2 | 380.0 | 7.9 | 406.8 | 8.0 |

From the spectra in Figure 2, by estimating the relative area under the peaks (1T/2H) associated to the different phases, it is possible to have an indication of their relative importance. We have calculated such a ratio (1T/2H) to 0.4 %, 12.1% and 15.4 % for $MoS_2$ films grown on STO, SiC and $Al_2O_3$ respectively. The results obtained by the Raman Spectroscopy measurements seem to indicate a substantial role played by the substrate in determining the heterogeneous phase formation in the $MoS_2$ films. In particular, these results point towards possible substrate induced effects mainly located at the basal planes. It is, therefore, reasonable to infer that $MoS_2$ films contain localized regions where the substrate promotes the coexistence of 1T and 2H domains.

In addition, the $E_{2g}$/$A_{1g}$ intensity ratio is larger in the case of $Al_2O_3$ substrate. Despite $E_{2g}$ and $A_{1g}$ are the characteristic vibrational modes of the 2H phase, their intensity ratio changes among samples reflecting modifications in interlayer coupling and electronic screening that correlate with an increased contribution of 1T-related Raman features. The full width half maximum (FWHM) of the $E_{2g}$ and $A_{1g}$ modes is smallest for the film grown on $Al_2O_3$, increases slightly for STO, and is largest for SiC, as reported in Table 1. This behaviour confirms that the vertical stacking of $MoS_2$ layers is most coherent on $Al_2O_3$ and is less correlated on STO and SiC, in agreement with the interpretation that the higher $E_{2g}$/$A_{1g}$ ratio on $Al_2O_3$ is related to an enhanced electronic screening from the 1T phase rather than structural disorder.

Our XRD and Raman results indicate that among the three substrates investigated, *c*-plane $Al_2O_3$ favours a higher fraction of the 1T phase in $MoS_2$ films, likely due to the combined effects of substrate-induced stacking correlations and the electronic environment at the surface plane.

To confirm these findings and to analyse the influence of the 1T phase on the electrochemical properties of $MoS_2$ films, we started by performing LSV measurements. LSV is widely exploited and considered a reliable method to evaluate the electrocatalytic performance of an electrode over a broad range of overpotentials. In analysing the LSV data, we point out that the HER kinetics are very complex and generally dependent on the electrochemical potential and on the pH value. The formation route of $H_2$ relies on the hydrogen coverage on the electrode surface and, in the acidic media, generally occurs with the simultaneous Tafel and Heyrovsky reactions. The first step, common to both these reactions, always comprises the adsorption of a hydrogen atom (H*) on an empty site (*) at the cathode (catalyst) surface through the combination of $H^+$ ions (protons) and electrons ($e^-$), and is known as the Volmer step ($H^+ + e^- + * \rightarrow H^*$). Once this step has been done, the $H_2$ formation in the Tafel process ($H^* + H^* \rightarrow H_2$) involves the combination of two adjacent adsorbed hydrogen atoms. Te other possible reaction to form $H_2$ is the Heyrovsky process ($H^* + H^+ + e^- + * \rightarrow H_2$), in which the proton ($H^+$) and the electron ($e^-$) directly adsorb on the adsorbed hydrogen atom (H*) on the electrode surface and form $H_2$. The theoretical slopes of Volmer, Heyrovsky and Tafel processes, reported in literature, are 118 mV/dec, 40 mV/dec and 30 mV/dec, respectively[45,46].

In Figure 3 we show the results of the LSV measurements (3a) performed on the $MoS_2$ films deposited on different substrates, along with the values of the slope of the associated j-V curves (3b), the so called Tafel slope. Following the Butler-Volmer equation [46]

$$j = j_0 \left( e^{-\alpha f \eta} - e^{(1-\alpha) f \eta} \right) \qquad \text{Eq. 1}$$

where α is the transfer coefficient, $f$ =F/RT and $\eta$ the overpotential, the linear form of the Tafel equation is given by η=a+b·log|j|, which provides an estimation of the electrocatalyst's performance. Important parameters related to the HER activity extracted from Eq.1 include the onset overpotential (η), i.e. the potential required to start the HER, the overpotential ($\eta_{10}$) at 10mA/cm$^2$, the exchange current density $j_0$ which is obtained by extrapolating the Tafel equation to the intercept of overpotential at 0V ( η= 0) and the Tafel slope of the linear portion of the Tafel equation. In Table 2, we report the measured values of these parameters for all the investigated films. The measured values of the Tafel slope reported in Table 2, relate well with the ones observed in previous work on 1T-2H $MoS_2$ films (around 72-77 mV/dec)[25]. The value of the onset overpotential at 1mA/cm$^2$ for the film deposited on $Al_2O_3$ is lower compared to those observed for the films on SiC and STO. The $\eta_{10}$ potential, required to obtain an output current density of 10mA/cm$^2$, has also a minimum value for the film deposited on the $Al_2O_3$ substrate. The results obtained by the LSV measurements confirm, the influence of the substrate on the electrocatalytic activity of $MoS_2$ films. In particular, we have observed the best electrocatalytic performances in the case of the films deposited on $Al_2O_3$. Based on the XRD and Raman analyses, this enhancement could be attributed to the formation or stabilization of the 1T phase.

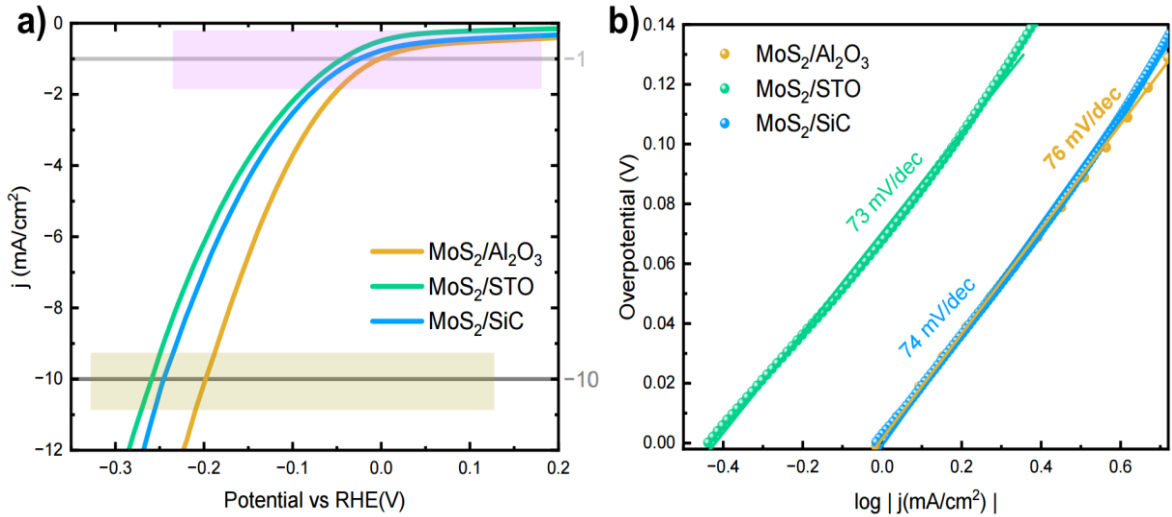

Figure 3. (a) Electrocatalytic performance evaluation of MoS$_2$ on different substrates, (b) Tafel slope estimation from corresponding j-V curves.

As a further check of the influence of the substrates on the electrochemical performances of MoS$_2$ films, we performed EIS measurements.

Table 2 Assessment of electrocatalytic HER performance derived from LSV data

| Film Assessment | MoS$_2$/Al$_2$O$_3$ | MoS$_2$/STO | MoS$_2$/SiC |
|---|---|---|---|
| Onset overpotential (mV) | 9 | 43 | 24 |
| Overpotential ($\eta$10) @10mA/cm$^2$ (mV) | 200.0 | 258.4 | 245.4 |
| Exchange current density, J$_0$ (A/cm$^2$) | 9.8E-4 | 3.8E-4 | 9.6E-4 |
| Tafel slope (mV/dec) | 76 | 73 | 74 |

EIS is a valuable, non-destructive technique for studying the interface between a heterogeneous catalyst and the electrolyte. The equivalent circuit model (ECM) is widely used to interpret EIS results for electrode processes involving adsorbed intermediates and to simulate HER behaviour in acidic and alkaline media.[43] In particular, it allows to provide insights into parameters like the double-layer capacitance (C$_{dl}$) and the polarisation capacitance (C$_a$), which are known to be linked to the electrochemical and electrocatalytically active surface area. The results are reported in Figure 4. We collected impedance spectra at progressively more negative overpotentials (0, –0.1, –0.20, –0.31V vs RHE) to supplement Tafel analysis and better understand the HER kinetics. The Kramers–Kronig consistency test confirmed that the system satisfied linearity, causality and stability under high potential. As shown in Figure 4 (a-c), the system's impedance dropped significantly as the overpotential increased, with depressed semicircles reflecting the intensification of the HER activity.

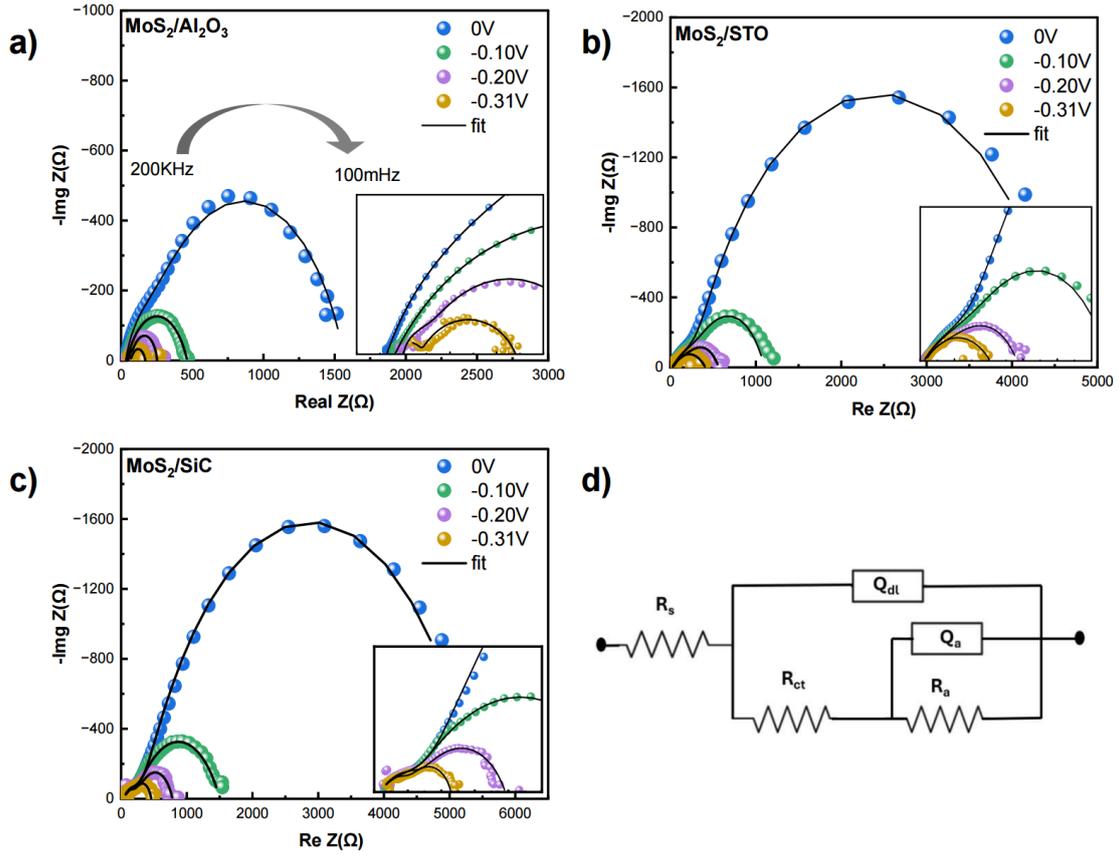

Figure 4. Electrochemical response of MoS$_2$ as function of increasing cathodic potential on different substrates (a) MoS$_2$ on Al$_2$O$_3$, (b) MoS$_2$ on STO, (c) MoS$_2$ on SiC, (d) Equivalent circuit model for fitting of impedance spectra.

LSV measurements revealed the beginning of the hydrogen evolution below –0.1V. Then, from -0.1V to -0.3V overpotentials, two semicircles became increasingly evident in the Nyquist plots, indicating separate electrochemical processes occurring at the film-substrate interface. The Nyquist plots were analysed using the circuit model shown in Figure 4(d), which is generally associated to the complex Volmer–Heyrovsky reaction pathway[47]. In this model $R_S$ is the electrical resistance associated to the electrolyte solution, $R_{ct}$ represents the electrical resistance associated with the charge transfer process occurring at the interface between the electrode and the electrolyte and $R_a$ is the polarisation resistance. In addition, $Q_{dl}$ and $Q_a$ are the so-called constant phase elements accounting for the surface's non-uniformity, associated to the $C_{dl}$ double layer capacitance and to the $C_a$ pseudo capacitance arising from hydrogen, calculated through the following equation proposed by Brug et al.[48]

$$C_{dl} = Q_{dl}^{1/n} \left(\frac{1}{R_s} + \frac{1}{R_{ct}}\right)^{1-\frac{1}{n}} \quad \text{Eq. 2}$$

$$C_a = Q_a^{1/n} \left(\frac{1}{R_s+R_{ct}} + \frac{1}{R_a}\right)^{1-\frac{1}{n}} \quad \text{Eq. 3}$$

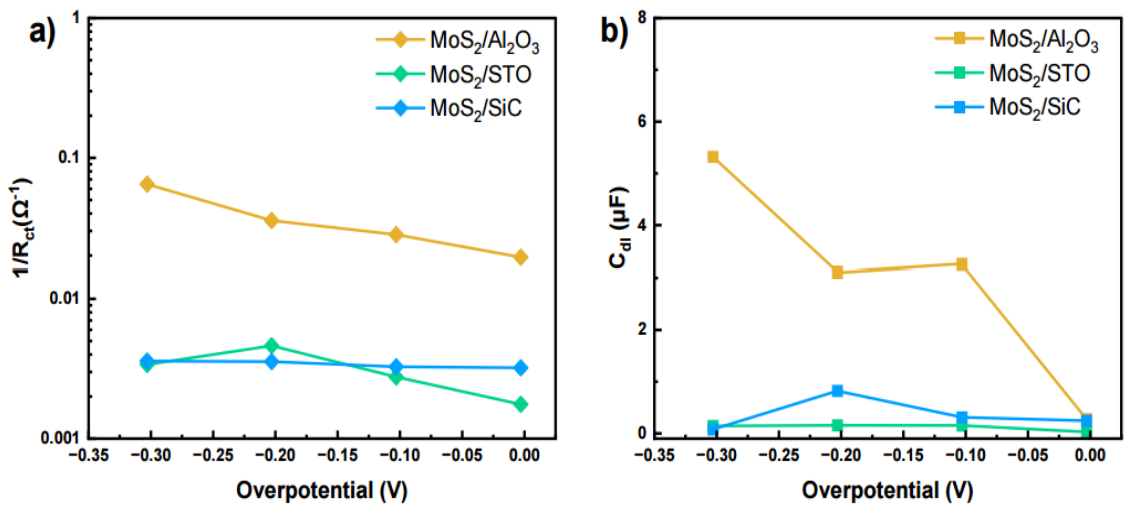

Figure 5. (a) $1/R_{ct}$ plot as a function of cathodic overpotential that corresponds to the reaction rate of $MoS_2$ film on different substrates (b) calculated double layer capacitance $C_{dl}$ that is correlated to electrochemical area during HER

As shown by the continuous lines in Figure 4(a-c), the model in Figure 4d fits very well all the experimental EIS spectra allowing the estimation of all the involved parameters. The measured double-layer capacitance ($C_{dl}$) is directly proportional to the electrode surface area exposed to the actual electrochemical activity[49,50]. As shown in Figure 5b, in the case of $MoS_2$ films on SiC and STO, $C_{dl}$ values are in the range of 0.03-0.15 µF and do not importantly vary with the overpotentials. On the other hand, the films on $Al_2O_3$ show higher $C_{dl}$ values consistently increasing with increasing overpotentials, up to 5.3 µF at -0.3V. This behaviour can be drawn back to the differences in the charge transfer mechanism at work in the investigated samples. In fact, from the curves in Figure 5a, the inverse of the charge transfer resistance ($1/R_{ct}$) measured for the $MoS_2$ films on $Al_2O_3$, is almost one order of magnitude higher than the values observed for the films deposited on SiC and STO. Assuming a simplified Volmer–Heyrovsky mechanism, $1/R_{ct}$ should be directly proportional to the HER reaction rate. The low frequency semicircles observed in Figure 4(a-c) are generally assigned to the $R_a$ and $Q_a$ elements of the equivalent circuit in Figure 4d.

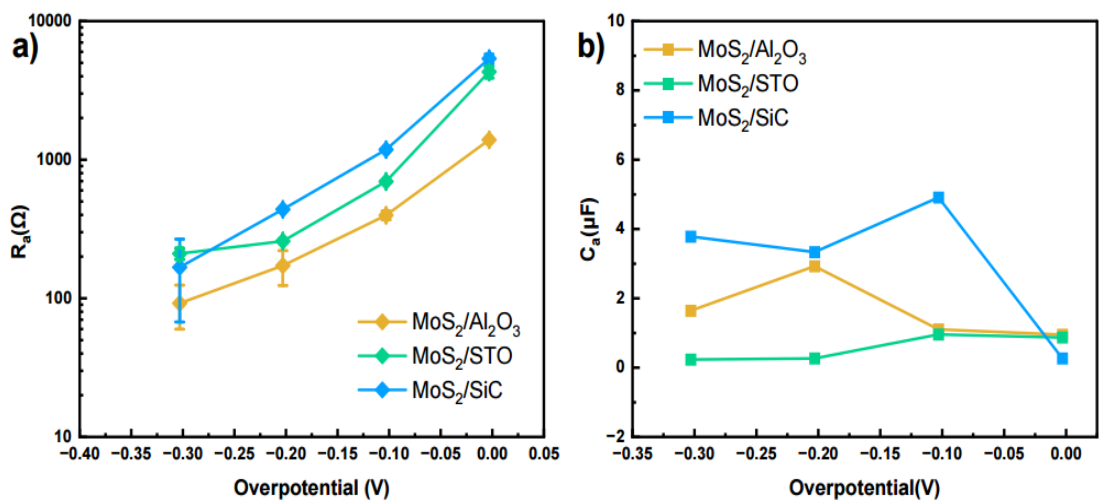

Figure 6. Low frequency values as a function of overpotential of (a) $R_a$ adsorption/desorption resistance and (b) $C_a$, adsorption/desorption capacitance.

In Figure 6(a-b) the measured values for these parameters obtained for the films deposited on different substrates, are shown. Both the adsorption/desorption capacitance $C_a$ and the adsorption resistance $R_a$ are sensitive to changes in overpotential. $C_a$ is known to be a valuable parameter for describing the hydrogen adsorption/desorption process[51]. In particular, $C_a$ values are related to the modulation of surface coverage by adsorbed hydrogen ($H_{ads}$) and to the number of active sites (Supplementary Information (SI section 2)). For samples on SiC and $Al_2O_3$, $C_a$ values show a slight tendency to increase with the increasing overpotential. Interestingly, those for sample on SiC are higher than those observed for films on $Al_2O_3$, probably corresponding to a more favourable adsorption/desorption mechanism. Moreover, $R_a$ is related to the electrical resistance associated with the hydrogen adsorption/desorption process. The $R_a$ values, shown in Figure 6a, exhibit a very rapid decline with increasing negative overpotential, being the most significantly affected parameter in this analysis and decreasing from around 5 k$\Omega$ at 0 V to 92 $\Omega$ at -0.3 V. The data in Figure 6a, do not show large variation at a given overpotential as a function of the different substrates, with the values obtained for the samples on $Al_2O_3$ being always the lowest.

The EIS data give important insights into the electrochemical activity of our $MoS_2$ films. In particular they allow to address the importance of substrate choice in optimising the charge transfer mechanism. From the curves in Figures 5 and 6 and from the ECM analysis, it is evident that the films on $Al_2O_3$ present the highest values of $1/R_{ct}$ and $C_{dl}$ at all the investigated overpotentials, with a strong dependence. Although the low frequency data indicate a higher number of active sites (SI section 2) in the sample on SiC compared to that on $Al_2O_3$, this difference seems to be less significant when weighed against the improved kinetics associated with the superior $1/R_{ct}$ and $C_{dl}$ values observed for $Al_2O_3$, in light of the overall electrochemical performance. Along with the LSV measurements, the EIS analysis allows to confirm the importance of the choice of suitable substrates to maximize the electrocatalytic properties of $MoS_2$ films, addressing the fast charge transfer across the electrocatalyst–electrolyte interface as the most critical mechanism, at least among the substrates investigated in this research. The cross-check of the LSV and EIS measurements with the XRD and the Raman results strongly suggests that the presence of the 1T phase is the relevant parameter to facilitate efficient electron transport and to promote the HER kinetics.

## Experimental
### $MoS_2$ Film deposition

$MoS_2$ thin films were synthesized via the pulsed laser deposition (PLD) technique using a KrF excimer laser ($\lambda$=248 nm) at the NFFA facility[52]. The films were deposited from a stoichiometric $MoS_2$ target under identical growth conditions on three different single-crystal substrates: $Al_2O_3$ (0001) (*c*-plane), $SrTiO_3$ (111), and 6H–SiC (0001). The laser energy density was maintained at approximately 2–2.5 J/cm$^2$ with a repetition rate of 1 Hz. The deposition was carried out under ultra-high vacuum conditions at a substrate temperature of $\approx$700°C. The growth rate was approximately 0.06 Å per laser shot. The film thickness was about 60nm on $Al_2O_3$ (0001) and 80nm for both $SrTiO_3$ and SiC.

## X-ray diffraction (XRD) Characterization

XRD measurements were carried out using a PANalytical EMPYREAN diffractometer equipped with a double-bounce Ge (220) monochromator on the incident beam and a point detector on the diffracted side.

## Raman Spectroscopy

The structure of the film is investigated through a micro-Raman InVia Renishaw spectrometer (1800 lines/mm grating, air-cooled double frequency He-Ne laser with λ=633 nm, power 6mW) with an air-cooled CCD camera and 50x VIS objectives Leica (NA=0.75) in N-plan.

## Electrochemical Characterization

All the samples were subjected to electrochemical measurements such as Linear sweep voltammetry (LSV), cyclic voltammetry (CV) and electrochemical impedance spectroscopy (EIS) performed in a 3-electrode standard cell setup using Zennium potentiostat/galvanostatic system. Linear sweep voltammetry (LSV) was performed at a scan rate of 10 mV/s, with the potential swept in the forward direction from 0.2 V to –0.8 V versus the Ag/AgCl reference ($E^0_{Ag/AgCl}$ =0.237 V) electrode (in 1M KCl) in 0.5 M $H_2SO_4$ (electrolyte), for hydrogen evolution reaction electrocatalysis studies. The potential window for CV is chosen from 0.25V to 0.42V with different scan rates ranging from 10 to 500 mV/s (10, 20, 50, 100, 200 and 500 mV/s). EIS measurements were conducted at progressively increasing potentials (0V to -0.31V vs RHE). The impedance spectra were acquired over 100 mHz to 200 kHz, using a perturbation amplitude of 10 mV. The system demonstrated satisfactory linearity, causality, and stability. The spectra were analysed using an equivalent circuit (ECM) model. The solution resistance (Rs) reported in Table S5s (see Supplementary Information), is extracted from the fitting and with the ohmic drop (iR-drop) 90% compensated during the experiment. The applied potentials were converted to values relative to the reversible hydrogen electrode (RHE) to determine the overpotential accurately:

$$E_{(RHE)} = E_{applied} + 0.059*pH + E^0_{Ag/AgCl}$$

## Conclusions

We have deposited by PLD thin films of $MoS_2$ on different substrates to investigate the role played by the substrate on the promotion of 1T and 2H different polymorphic structures and on the HER kinetics. Performing XRD, Raman, LSV and EIS measurements, we demonstrate that the strategic selection of substrate enables the growth of distinct $MoS_2$ polymorphs, resulting in tuneable film intrinsic properties and activity per site. Among the substrates here investigated, the films deposited on $Al_2O_3$ exhibit the highest electrocatalytic activity during the HER in acidic media. From the XRD and Raman analyses, this behaviour could be traced back to the substrate-induced modification of the 1T/2H phase ratio, which affects the electronic structure and facilitates more efficient charge transfer due to the metallic character and higher density of states of the 1T phase near the Fermi level, ultimately leading to optimized charge transfer resistance values. This study opens the way to further studies on the modulation of the HER catalytic activity of $MoS_2$ films by suitably choosing the substrate able to facilitate the HER during water splitting.

## Author contributions

H. Sami-Ur-Rehman: conceptualization, formal analysis, investigation, writing – original draft, and writing – review and editing, N. Coppola, P. Polverino, D. Sannino, H.-C. Neitzert, and S. Punathum-Chalil: formal analysis and investigation. A. Singh, and S. K. Chaluvadi: formal analysis, investigation, writing – original draft. P. Orgiani, and P. Barone: conceptualization, formal analysis, investigation, writing – review and editing. A. Galdi, and C. Pianese: supervision, and writing – review and editing. C. Aruta, and L. Maritato: conceptualization, formal analysis, funding acquisition, investigation, project administration, supervision, writing – original draft, and writing – review and editing.

## Conflicts of interest

There are no conflicts to declare.

## Data availability

The data supporting this article have been included as part of the Supplementary Information. Supplementary information: calculation of the mismatch (Figure S1 and Table S1), estimation of the number of active sites (Table S2), cyclic voltammetry and double layer capacitance measurements (Figure S2), double layer capacitance (Table S3) and absorption resistance values (Table S4) estimated from equivalent circuit model, solution resistance from impedance spectra (Table S5).

## Acknowledgements

We acknowledge support from the Italian Ministry of Research under the PRIN 2022 Grant No 202228P42F with title "Transition metal dichalcogenide thin films for hydrogen generation" PE3 funded by PNRR Mission 4 Istruzione e Ricerca - Componente C2 - Investimento 1.1, Fondo per il Programma Nazionale di Ricerca e Progetti di Rilevante Interesse Nazionale PRIN 2022 – CUP B53D23003800006.
Research at SPIN-CNR was also supported by the project ECS00000024 "Ecosistemi dell'Innovazione"— Rome Technopole of the Italian Ministry of University and Research, public call n. 3277, PNRR— Mission 4, Component 2, Investment 1.5, financed by the European Union, Next Generation EU.

# Supplementary Information

# Substrate induced optimization of the Electrocatalytic Hydrogen Evolution Reaction (HER) performances of MoS$_2$ thin films

Hafiz Sami-Ur-Rehman,*[a] Arpana Singh,[a] Nunzia Coppola,[a] Pierpaolo Polverino,[a] Sandeep Kumar Chaluvadi,[b] Shyni Punathum-Chalil,[bc] Heinrich-Christoph Neitzert,[a] Diana Sannino,[a] Pasquale Orgiani,[bd] Alice Galdi,[a] Cesare Pianese,[a] Paolo Barone,[e] Carmela Aruta,[e] Luigi Maritato[a]

[a] *Dipartimento di Ingegneria Industriale-DIIN, Università Degli Studi di Salerno, 84084 Fisciano (SA), Italy.*
[b] *CNR-IOM, Strada Statale 14 Km 163.5, 34149 Basovizza, Trieste, Italy.*
[c] *International Centre for Theoretical Physics (ICTP), Str. Costiera 11, I-34151 Trieste, Italy.*
[d] *AREA Science Park, Padriciano 99, I-34149 Trieste, Italy.*
[e] *CNR-SPIN, Via del Fosso del Cavaliere 100, 00133, Roma, Italy.*
* Email: hsamiurrehman@unisa.it

**Section 1**

In the case of substrate mismatch, due to the lack of epitaxy, we can only account for local distortions within individual domains. In Fig. S1, we attempted to evaluate the possible in-plane mismatch between the MoS$_2$ film and the different substrates in the most favourable conditions. It follows that the best matching can be obtained by considering twice the in-plane lattice parameter of Al$_2$O$_3$ and three times the in-plane parameter for the MoS$_2$ film and the SiC substrate, while in the case of SrTiO$_3$ (STO) we must consider the triangle reported in Figure 1 obtained from the [110] and double [11-2] crystal directions. The mismatch that we obtain are reported in Table S1.

First, it can be inferred that Al$_2$O$_3$ and SiC exhibits – in absolute terms - the largest and the lowest lattice mismatch with both the 2H and 1T phases. As a consequence, MoS$_2$ films grown on Al$_2$O$_3$ generally show narrow and more intense diffraction peaks which can be correlated to higher structural order. On the contrary, MoS$_2$ films grown on SiC substrate appear more disordered as evidenced by the XRD measurement, which displays a broader and weaker diffraction peak compared to the previous case. Moreover, the expansion of the out-of-plane *c* lattice parameter in MoS$_2$ films grown on SiC is compatible with an in-plane compressive strain induced by the substrate.

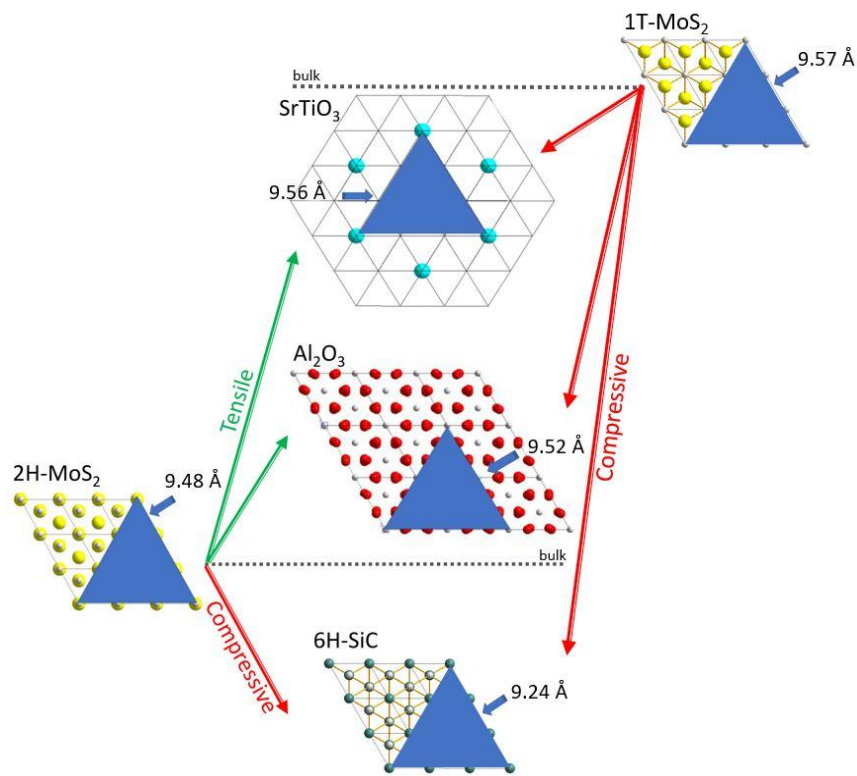

Figure S1 In-plane mismatch between the 1T and 2H phases of MoS$_2$ film and the STO, Al$_2$O$_3$ and SiC substrates.

In contrast, the case of MoS$_2$ films grown on STO appears to be more complex. If we focus on the average mismatch, MoS$_2$ films grown on STO substrate should experience a larger tensile strain with respect to the case of Al$_2$O$_3$. As a consequence, a further shrinking of the MoS$_2$ out-of-plane parameter is expected when compared with films grown on sapphire. However, this is not happening thus raising concerns regarding a purely elastic deformation of the MoS$_2$ crystallographic structure induced by substrate-related strain, which is currently under investigation.

Table S 1 Mismatch calculated considering the in-plane orientation reported in Fig. S1

| Substrate | Mismatch with 2H | Mismatch with 1T | average mismatch |
|---|---|---|---|
| STO | + 0.6 % | - 0.1 % | + 0.2 % |
| Al$_2$O$_3$ | + 0.1 % | - 0.5 % | - 0.2 % |
| SiC | - 2.8% | - 3.4 % | - 3.1 % |

Nevertheless, the use of these three-substrate allowed us to deform the MoS$_2$ crystal structure and to probe electrochemical properties as a function of these last.

**Section 2**

Electrochemical impedance spectroscopy (EIS) can be also employed to estimate the number of active sites on $MoS_2$ electrocatalysts. Theoretically, the surface coverage of adsorbed hydrogen is expected to approach unity ($\Theta \approx 1$) at sufficiently negative overpotentials [1].

In the present study, a similar analytical approach was adopted to estimate the number of active sites on $MoS_2$ films deposited on different substrates. Importantly, both the physical and mathematical estimation of active site density depends on the applied overpotential. Hence, the calculations were conducted using the most cathodic potential explored in this study, −0.2 V.

Several assumptions were made to simplify the analysis: hydrogen adsorption proceeds via a one-electron transfer step, each active site binds one proton ($H^+$), and $\Theta$ is approximately 1. Under these conditions, the total charge ($Q_t$) associated with a full monolayer of adsorbed hydrogen can be estimated by:

$$Q_t = C_a \times \eta_{-0.2}$$

where $C_a$ is the adsorption capacitance calculated from equation 3 and reported in Table S4 and $\eta$ is the overpotential −0.2 V.

The number of active sites ($N_s$) can then be determined using:

$$N_S = \frac{Q_t \times N_a}{nF}$$

Here, $N_a$ is Avogadro's number, $n$ is the number of electrons involved per reaction ($n = 1$), and $F$ is Faraday's constant (96485 C/mol).

As shown in Figure 6b, at sufficiently negative overpotentials the observed pseudo-capacitance reflects the degree of surface coverage during hydrogen adsorption and desorption on $MoS_2$ films grown on SiC and $Al_2O_3$. This behavior is typically associated with the density of accessible active sites on the catalyst surface. Based on the linear sweep voltammetry (LSV) data, one might expect that the $Al_2O_3$ supported films exhibit the highest density of active sites. Interestingly, our results show the opposite trend, with SiC substrates yielding greater site availability compared to the other films. Since HER is fundamentally governed by the adsorption/desorption of hydrogen intermediates, this difference is of critical importance. Although the SiC-supported films display higher surface coverage, their electrocatalytic performance is limited by a relatively larger charge transfer resistance ($R_{ct}$), which suppresses overall hydrogen evolution reaction (HER) kinetics. In contrast, for $SrTiO_3$ (STO), both surface coverage and charge transport are less favourable, resulting in the poorest catalytic activity among the substrates tested. Taken together, these complementary analyses highlight that substrate choice not only dictates interfacial charge transfer but also modulates surface coverage, thereby playing a key role in determining HER activity.

Table S2. Number of sites available for HER on different substrate

| Total Active site | MoS$_2$/Al$_2$O$_3$ | MoS$_2$/STO | MoS$_2$/SiC |
|---|---|---|---|
| Na x 10$^{13}$ | 7.4 | 0.7 | 8.5 |

The differences in HER performance observed for MoS$_2$ films grown on different substrates cannot be explained solely by variations in surface morphology. In particular, Al$_2$O$_3$ substrates appear to modulate the intrinsic catalytic properties of MoS$_2$ by influencing the interfacial charge transfer resistance. Although some uncertainties remain in the quantitative analysis, the overall trend consistently demonstrates Al$_2$O$_3$ substrates enhance the catalytic performance of MoS$_2$ films.

**Section 3**

The cyclic voltammetry (CV) measurements were conducted within a potential window of 0.25–0.42 V vs Ag/AgCl, as shown in Figure S2 (a-c). The current density difference, defined as $\Delta j = (j_a - j_c)/2$ at 0.35 V vs Ag/AgCl, was plotted as a function of scan rate (Figure S2 (d)) to estimate the double-layer capacitance ($C_{dl}$). The results show that MoS$_2$ films on SiC substrates exhibit a larger electrochemically accessible surface area, which is advantageous for charge storage applications. However, the increased charge transfer resistance observed in this case reduces the overall electrocatalytic activity.

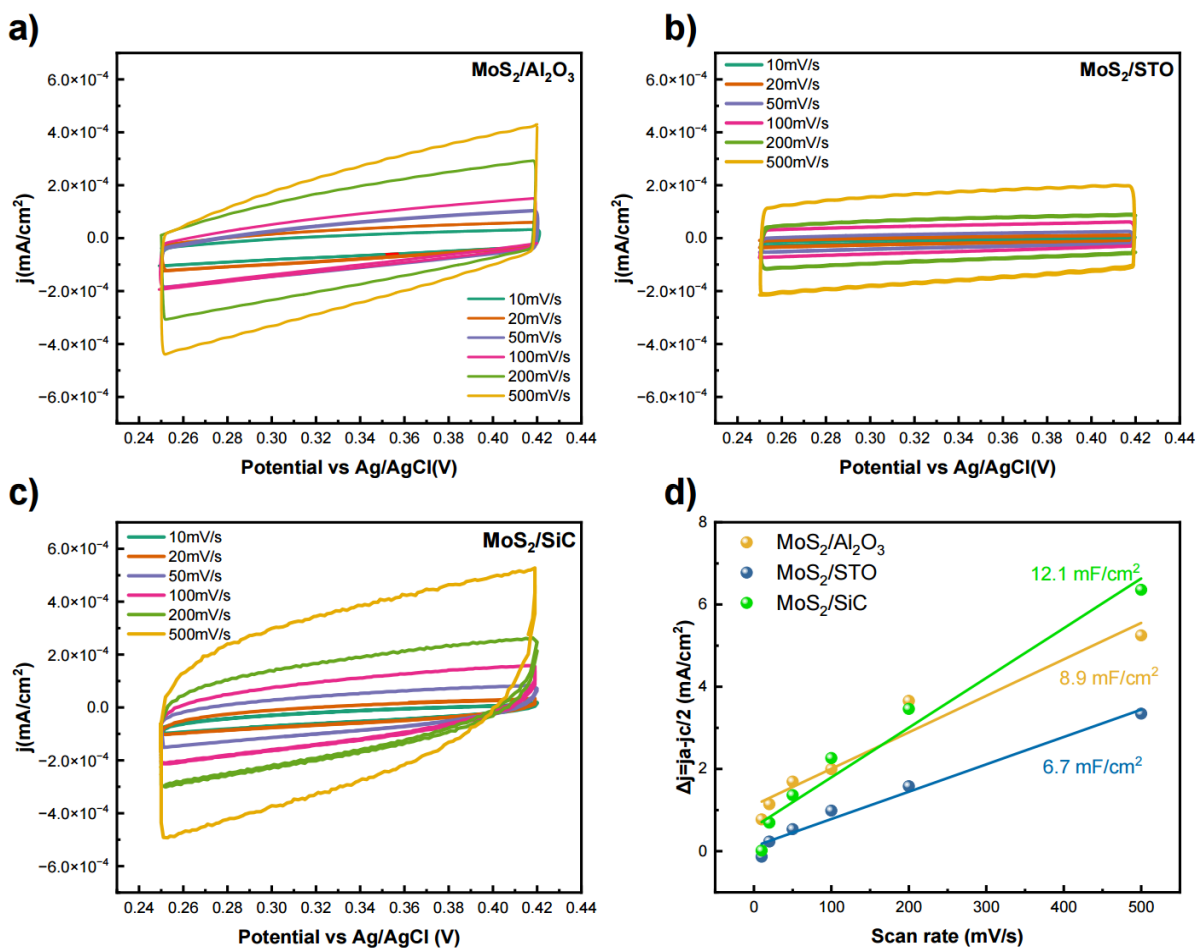

Figure S2. CV scan as function of scan rate of MoS$_2$ film on different substrate (a) Al$_2$O$_3$, (b) STO, (c) SiC and (d) current density difference and double layer capacitance estimations from panels a-c.

The double-layer capacitance reported in Figure S2(d) is determined from the non-faradaic region of the CV response of Figure S2 (a-c). It provides a reliable estimate of the electrochemically active surface area (ECSA). This analysis supports the earlier discussion on surface coverage presented in Section 2.

Table S3. Double layer capacitance (C$_{dl}$) estimated from equivalent circuit model reported in Figure 4 d

| Vs RHE Overpotential (V) | MoS$_2$/Al$_2$O$_3$ | | MoS$_2$/STO | | MoS$_2$/SiC | |
|---|---|---|---|---|---|---|
| | C$_{dl}$ (µF) | Error ($\chi^2$) | C$_{dl}$ (µF) | Error ($\chi^2$) | C$_{dl}$ (µF) | Error ($\chi^2$) |
| 0.00 | 0.28 | 0.09 | 0.03 | 0.07 | 0.25 | 0.05 |
| -0.10 | 3.20 | 0.01 | 0.16 | 0.05 | 0.31 | 0.05 |
| -0.20 | 3.10 | 0.08 | 0.16 | 0.06 | 0.82 | 0.12 |
| -0.31 | 5.32 | 0.11 | 0.15 | 0.13 | 0.08 | 0.05 |

Table S4. Values of the absorption resistance ($R_a$) and absorption/desorption capacitance ($C_a$) obtained from equivalent circuit model reported in Figure 4d.

| η /Vs RHE (V) | MoS$_2$/Al$_2$O$_3$ | | | MoS$_2$/STO | | | MoS$_2$/SiC | | |
|---|---|---|---|---|---|---|---|---|---|
| | Ra (Ω) | Ca (μF) | Error ($\chi^2$) | Ra (Ω) | Ca (μF) | Error ($\chi^2$) | Ra (Ω) | Ca (μF) | Error ($\chi^2$) |
| 0.00 | 1391.2 | 0.95 | 0.09 | 4305.9 | 0.87 | 0.07 | 5362.3 | 0.27 | 0.05 |
| -0.10 | 397.8 | 1.10 | 0.01 | 696.05 | 0.96 | 0.05 | 1183.9 | 4.91 | 0.05 |
| -0.20 | 172.1 | 2.93 | 0.07 | 258.6 | 0.57 | 0.06 | 437.88 | 3.33 | 0.12 |
| -0.31 | 92.2 | 1.64 | 0.11 | 210.24 | 0.23 | 0.13 | 167.4 | 3.78 | 0.05 |

Table S5. The solution resistance ($R_s$) calculated from the impedance spectra as a function of cathodic overpotential

| Vs RHE Overpotential (V) | Solution Resistance ($R_s$)/(Ω) | | |
|---|---|---|---|
| | MoS$_2$/Al$_2$O$_3$ | MoS$_2$/STO | MoS$_2$/SiC |
| 0.00 | 30.62 | 21,47 | 47.35 |
| -0.10 | 37.11 | 18.61 | 43.68 |
| -0.20 | 47.68 | 21.40 | 43.70 |
| -0.31 | 55.11 | 30.11 | 41.50 |